\documentclass[a4paper]{article}

\usepackage[english]{babel}
\usepackage[]{graphicx}
\usepackage[latin1]{inputenc}
\usepackage[T1]{fontenc}
\usepackage{charter}
\usepackage{tabularx}
\usepackage{lscape}

\usepackage[margin=1.2in]{geometry}

\usepackage{amssymb}
\usepackage{pdfpages}

\usepackage[colorlinks,linkcolor=blue,citecolor=blue,urlcolor=blue]{hyperref}

\usepackage[]{listings}
\usepackage{enumitem}

\newcommand{\code}[1]{\textsf{#1}}

\title{What Should Developers Be Aware Of? An Empirical Study on the Directives of API Documentation}

\author{
Martin Monperrus$^{1,2}$
\and 
Michael Eichberg$^1$
\and 
Elif Tekes$^1$
\and 
Mira Mezini$^1$
}

\date{}

\usepackage{textpos}

\begin{document}

\maketitle

\begin{textblock*}{\textwidth}(0cm,21cm)
\noindent\small In: Empirical Software Engineering, Springer, Online Edition, 2011. DOI: 10.1007/s10664-011-9186-4
\end{textblock*} 

\footnotetext[1]{Technische Universit\"at Darmstadt, Germany}
\footnotetext[2]{University of Lille, France}

\begin{abstract}
Application Programming Interfaces (API) are exposed to developers in order to reuse software libraries.
API directives are natural-language statements in API documentation that make developers aware of constraints and guidelines related to the usage of an API. 
This paper presents the design and the results of an empirical study on the directives of API documentation of object-oriented libraries.
Its main contribution is to propose and extensively discuss a taxonomy of 23 kinds of API directives.
\end{abstract}

\section{Introduction}

\begin{quotation}
``Reuse is something that is far easier to say than
to do. Doing it requires both good design and
very good documentation. Even when we see
good design, which is still infrequently, we won't
see the components reused without good documentation.'' 
D. L. Parnas, cited by Fred Brooks, The  mythical man-month: Essays on software engineering \cite[page 224 (1995 edition)]{Brooks1975}
\end{quotation}

Developers of reusable software elements --- such as software libraries --- generally have the responsibility to provide high-quality and comprehensive documentation \cite{kramer1999api}.
As the above quotation shows, Parnas goes even further and states that reuse is not possible without good documentation.
However, the research on library documentation is in stark contrast to its importance \cite{Robillard2009}.
Actually, it is required to have a clear understanding of what documentation contains in order to build a wide range of software engineering tools. For example, to find inconsistencies between documentation and code or to automatically infer missing yet important documentation.
To improve the understanding of source code documentation, we did
 an extensive case study of existing library documentation.

We consider a \emph{library} as the implementation of some functionality that is meant to be reused by other developers.
In case of object-oriented programming languages, such as Java and C\#, a library consists of a coherent collection of classes.
In general, a library has two parts: the public interface and the private implementation.
The former contains software elements (e.g. classes and methods) that other developers are allowed and expected to use; this part is called the \emph{application programming interface} or API.
The latter implements the library's functionality, but is usually not exposed to the library's users and is not further considered in this paper.
For instance, the Java Development Kit (JDK) is a multipurpose library that provides as part of its API classes for input and output (IO) (e.g. \code{java.io.File}) and to build graphical user interfaces (using the \code{javax.swing} classes). However,  the details how the IO functionality is implemented on a given operating system and how the elements of the graphical user interface are created given a specific windowing toolkit are not exposed to the users of the library.

To foster reuse, \emph{application programming interfaces} (APIs) are delivered with some documentation which --- in its broadest sense --- explains how to use the library. 
Accordingly, the scope of API documentation is very large: it spans different kinds of information and also has to satisfy different kinds of audiences.
For instance, API documentation may contain typical usage scenarios, code snippets, design rationales, performance discussions, and contracts.
The scope of the targeted developers ranges from complete novices who have never used the API to experts who are able to tweak every single piece of the API.

In this paper, we concentrate on the parts of an API's documentation that describe directives \cite{Dekel2009a,Bruch2010}. API directives are natural-language statements that make developers aware of constraints and guidelines related to the usage of an API. 
Directives are manifold and are related to, e.g., development issues, performance questions or software evolution. In other terms, pieces of an API's documentation that describe how to correctly and optimally use a library are API directives.
This definition encompasses Dekel's \cite{Dekel2009a} and Bruch's \cite{Bruch2010} definition of API directive, but is a bit broader to provide enough leeway to accommodate natural language subtleties.

Next, we present three different directives from three different software packages to sketch the scope of API directives.

The documentation of the constructor of the standard Java class \code{java.se\-cu\-rity.Time\-stamp} states that \emph{``@param timestamp is the timestamp's date and time. It must not be null''}. In this case the directive is that the parameter must not be null. In general, if developers are not aware of this directive and pass a \code{null} value to the corresponding method, an exception will probably be raised. Violations of such directives are usually detected in a timely manner, likely during unit testing and can easily be resolved.

The directive \emph{``HashedMap is not synchronized and is not thread-safe''} (in the API documentation of Apache Commons Collections) is related to the proper synchronization of the class if it is used in a multi-threaded program. A violation of this constraint will eventually result in synchronization related bugs that are often erratic and hard to locate and to resolve.

The documentation of the constructor \code{Item(parent,style)} of the class \code{org.ec\-lipse.swt.widgets.Item} states that \emph{``the style value [an integer] is either one of the style constants defined in class SWT which is applicable to instances of this class, or must be built by bitwise OR'ing together (that is, using the int "|" operator) two or more of those SWT style constants.''}.
If a developer now passes an integer value where certain bits are not taken into account by the implementation, the behavior of the method is left undefined.
This means that if the SWT designers later on decide to use further bits, the behavior of the client program may change and behave unexpectedly.

To our knowledge, the research on the nature of API directives is very limited. 
Dekel \cite{Dekel2009a}
mentions them 
and some books on API design shallowly touch the topic (e.g. \cite{tulach2008practical}), but a thorough analysis has not been carried out so far.
To improve our understanding of API directives, we have designed and performed an empirical study that explores directives of large and widely used APIs. 
We (the authors) have read more than 6000 pieces of Java API documentation. For each of them, we tried to understand  whether it is related to a directive and, if it is a directive, how it can be classified.
The case-study protocol and the results are made public, to enable other researchers to conduct replication studies.

The contributions of this paper are:
\begin{itemize}
\item a comprehensive taxonomy of API directives that are present in API documentation of object-oriented libraries written in Java  (Figure \ref{fig:taxonomy}). Each directive kind of this taxonomy is discussed in depth and is supported by examples 
from the examined APIs.
Furthermore, for each directive kind we give a first estimation of its abundance to indicate how valuable it would be to devise automated approaches to support its specification, analysis or checking.
\item a set of empirical results on the links between words in an API's documentation and their importance with respect to pinpointing to API directives (Table \ref{tab:precision}). 
The collected data is publicly available as electronic supplementary material as well as in \cite{paperdata}. Subsequent work may use it to calibrate documentation tools that use natural language processing, for instance to highlight the most important phrases according to a predicted likelihood.
\end{itemize}

The remainder of this paper is structured as follows.
Section \ref{related-work} presents the related work and 
Section \ref{methodology} describes the setup of our empirical study. The  taxonomy of directive kinds is given in Section \ref{list-of-directive-kinds}.
Section \ref{discussion} further discusses our findings. Section \ref{conclusion} concludes this paper.

\section{Related Work} 
\label{related-work}

In this section, we discuss other research related to software documentation. 
We first show that not all code comments are API documentation (\ref{rw-code-comments}),
then sketch the gap (\ref{rw-topdown}) between expert recommendations on how to write documentation and empirical guidelines, followed by the key difference (\ref{rw-tool-oriented}) between action and observational research.
We conclude by exposing other empirical studies (\ref{rw-empirical-studies}) on APIs and showing that the closest work (\ref{rw-closest-work}) -- to our knowledge -- is subject to improvement. 

\subsection{Code Comments}
\label{rw-code-comments}

Padioleau et al. \cite{padioleau2009listening} published an empirical study of comments in operating system code. They examined 1050 comments from three different operating system kernels and thereafter built a taxonomy of comments that has four axes: the content, the audience, the related source element and the evolution in time. 
Storey et al.~\cite{Storey2008} and Ying et al.~\cite{ying2005source} also did empirical work related to source code documentation.  
They focused on source code comments that represent so called task annotations (e.g. \emph{{TO}DO} or \emph{FIXME}).  
Jiang et al.~studied \cite{Jiang2006} the evolution of comments over time.
All these authors focused on code comments, which is different in nature from API documentation.

On the contrary, we are interested in API documentation that is written to enable the efficient use of libraries. This is very different from the goal of many source code comments found in operating system code which explain the code for the sake of maintenance and often provide insights about the development history.
Second, we focus on API comments of object-oriented libraries, whose scope  is significantly different from comments of procedural C code in the context of operating systems.
Similarly, API documentation has a completely different purpose when compared to task annotations. 

\subsection{Top-down Recommendations for API Documentation}
\label{rw-topdown}

Several authors defined what should be present in API documentation in a top-down way. I.e., they enumerated  lists of important pieces of documentation for each kind of API element (package, class, method, field).
For instance, Sun published a technical report entitled ``Requirements for Writing Java API Specifications'' \cite{Smith2003}.
One of the lead developers of the Eclipse platform\footnote{Eclipse is an extensible development environment primarily targeted for Java.}, Boris Bokowski also emphasized providing appropriate and precise content in API documentation \cite{Bokowski2006,Bokowski2008}.

Our work differs from this in that 1) we follow a bottom-up approach, i.e.\ we first did an empirical study of the documentation of high-quality libraries before formulating some guidelines and 2) we only focus on the part of API documentation related to directives and not on the whole API documentation.

\subsection{Tool-oriented Research}
\label{rw-tool-oriented}

Especially the recent work by Stylos et al. \cite{Tan2007,stylos2009jadeite,Stylos2009,Stylos2009a} explores various kinds of tool-support for software documentation.
For instance, Jadeite \cite{stylos2009jadeite} enriches Java API documentation with additional information aggregated from real users in order to help developers to find the right class or the right method.
Along the same line, Dekel and Herbsleb \cite{Dekel2009a} proposed an approach and a tool to improve the awareness of developers of directives.

On the contrary, our paper presents observational research. 
We do not devise or build a tool, but rather observe the reality and try to gain 
knowledge out of it.
However, this knowledge is meant to be used by future work in order to build advanced documentation tools.

\subsection{Empirical Studies on APIs}
\label{rw-empirical-studies}

Many empirical studies on APIs have been carried out.
For instance, authors explored the issue of API usability \cite{Clarke2004,Ellis2007,Stylos2008,Stylos2008a,Watson2009} with a focus on the API design.
Shi et al.~\cite{Shi2011}  concentrated on the evolution of API documentation and Parnin et al.~\cite{Parnin2011} on the power of the World Wide Web to find API documentation.

However, none of the previous studies analyzed the content and especially the directives of API documentation as we do in this paper.

\subsection{API Contracts and Directives}
\label{rw-closest-work}

Arnout and Meyer \cite{Arnout2003} showed in 2003 that hidden API contracts can be recovered from different sources of information, including API documentation. Their paper proves that hidden contracts actually exist by presenting some of them found in the .NET library. In contrast, our empirical study provides a comprehensive overview of those ``contracts'' that are explicitly stated. 

As part of his Ph.D. thesis Dekel \cite{Dekel2009} did a deep study related to API directives.
Both his approach and ours are empirical. 
However, three major differences exist compared to Dekel's seminal work. 
First, our study is supported by an empirical design that supports replication (see Section \ref{methodology}) while Dekel's study is more informal.
Second, our case-study protocol supports measuring the abundance of each kind of directives.
Third, our study is larger in scope: we tried to identify ``all'' kinds of directives.
In particular, we provide a thorough analysis of directives related to subclassing.
On the contrary, apart from protected versus private visibility directives, the directives related to subclassing and interface implementation were not in the scope of Dekel's work.

\section{Case Study Protocol}
\label{methodology}

In this section, we discuss the case study that we have carried out to answer the following research questions:
\emph{Which kinds of API directives exist?} 
\emph{How can API directives be classified?}
According to Easterbrook et al.~\cite{Easterbrook2007}, the first one is a ``description question'' and the second one a ``classification question'' and both types of questions can be answered with an exploratory case study.

The outcome of the case study is a taxonomy of API directives. By taxonomy, we mean a set of categories, logically organized in groups of related categories. The taxonomy is not meant to be exclusive, i.e. certain directives may belong to several categories. In the following, we use the term ``directive kind'' to refer to a category of API directives. Next, we first outline how we carried out the case study before we describe it in more detail.

\subsection{Case-study Protocol} 

To obtain a taxonomy of directives, we analyzed a large amount of API documentation in order to get a deep understanding of the nature of API directives. 
The overall process can be summarized as follows:
\begin{enumerate}
\item We first determined the corpus of API documentation to be analyzed (this step is further discussed in Section \ref{corpus}). For instance, the API of \texttt{java.io} is an element of the corpus that we have analyzed.
\item The second step was to identify the syntactic patterns that are likely to reveal directives. For instance, ``must'' is a syntactic pattern that may reveal a directive such as ``The method parameter must not be null.''
\item For each syntactic pattern, all occurrences of it are automatically identified in the analyzed API documentation. 
\item For each pair of syntactic pattern and API, we analyzed as many 
 occurrences of the given pattern as necessary to get statistically significant results. 
 Analyzing an occurrence means deciding whether it is a directive and -- if so -- whether it is a new kind of directive or an instance of an already identified directive kind.
\end{enumerate}

\begin{table}
\centering\begin{tabular}{|l|p{3cm}|p{3cm}|p{3cm}|}
\hline
       & JDK                   & JFace                 & Commons \newline Collections   \\
\hline 
Target Platform &     Java                    &          Java               &    Java   \\                  
\hline 
Development Model &     \small{Company-driven (Sun)}                    &   \small{Open-source and company-driven (IBM)}                      &    \small{Open-source and community-driven}  \\                  
\hline 
Purpose &     IO, Data structures, Concurrency, Security   &          User-interfaces              &    Data structures   \\

\hline
API Comment Location &                         &                         &       \\                  
~~~field &                        950 &                        648 &                        205 \\
~~~method &                       8\,361 &                       5\,918 &                       2\,586 \\
~~~class &                        671 &                        404 &                        263 \\
~~~interface &                        189 &                        264 &                         31 \\
~~~package &                         33 &                         38 &                         13 \\
           \hline
           \# API elements &                      10\,204 &                       7\,272 &                       3\,098 \\
           \hline
           \hline
                  \# Words &                    1\,055\,220 &                     264\,952 &                     114\,895 \\
\hline
\end{tabular}
\caption{Key figures of our API documentation corpus. The large majority of API comments is at the method level. There is a total of 20\,574 API documentation elements.
}
\label{descriptive-stats} 
\end{table}

\subsubsection{Determining an API Corpus}
\label{corpus}
Given the vast amount of available object-oriented libraries, it is necessary to limit the study to a set of representative libraries.

To identify possible candidate libraries, we first grossly characterized the libraries along the following three properties: ``\emph{the target platform}'', ``\emph{the development model}'' and ``\emph{the purpose}'' of the library.
Practically, all \emph{libraries have a specific purpose} and are developed to solve a set of well-defined, highly-related programming tasks. Nevertheless, the scope of libraries is 
 still very different. For instance, the scope is very narrow for a logging library or a library that implements graph algorithms such as JGraphT\footnote{see \url{http://jgrapht.sourceforge.net}}. But, in case of Java's \texttt{java.io} library the scope is broader. It provides functionality to read from/write to files and also sockets; i.e., it supports desktop as well as distributed applications.
Most \emph{libraries target a specific platform}, such as Java\footnote{see \url{http://www.oracle.com/technetwork/java/javase/index.html}} or .NET\footnote{see \url{http://www.microsoft.com/net/}}, and are highly dependent on it. Using such a library to develop an application that targets a different platform is often practically impossible. It would at least require the development of some special bindings to make the library useable on a different platform or even require a complete reimplementation of the library. 
The \emph{development model of libraries} range from completely free, community driven libraries (e.g. the C++ Boost library\footnote{see \url{http://www.boost.org/}}) over company-driven libraries that are open-source or at least free to use (e.g. the Microsoft Foundation Classes\footnote{see \url{http://msdn.microsoft.com/en-us/library/d06h2x6e(v=VS.100).aspx}}), to closed-source, commercial libraries.  

For our study, we consider the API documentation of packages from the official Java Development Kit, the Eclipse project, and from the Apache foundation.
For the JDK, we considered \code{java.lang}, \code{java.util}, \code{java.io}, \code{java.math}, \code{java.net}, \code{java.rmi}, \code{java.sql}, \code{java.security}, \code{java.text}, and \code{java.applet}. The API documentation was extracted from the source code of the JDK\footnote{ This version of JDK is available at \url{http://download.java.net/jdk6/6u21/promoted/b05/}}, version \code{jdk-6u21-ea-src-b05-jrl-29\_may\_2010.jar}. 
We also analyzed the API documentation of Eclipse's JFace library\footnote{see \url{http://wiki.eclipse.org/index.php/JFace}}, CVS version of Oct. 15 2009\footnote{\$ cvs -d :pserver:anonymous@dev.eclipse.org/cvsroot/eclipse/ co -D "15 Oct 2009" org.eclipse.jface} . JFace is a UI toolkit mainly used by the Eclipse IDE and respective plug-ins.
Finally, we also took into account the API of the Apache Commons Collections library\footnote{see \url{http://commons.apache.org/collections/}}, revision 1000473. The Commons Collections API provides general-purpose implementations of collection classes (e.g. List, Set, etc.). 

We selected these libraries because they have a wide scope and they are supported by large organizations. 
By choosing libraries that cover a wide range of use cases we try to alleviate the risk that the study is 
biased towards a particular 
domain. 
The decision to focus on APIs supported by large organizations is based on the evidential observation that the level and the quality of their documentation is often high. Large organizations typically have the financial means and also interest to support the documentation process; e.g. by employing technical writers.

Table \ref{descriptive-stats} shows the characteristics of the chosen APIs, and in particular the number of API documentation items per type of code element.
It shows the number of API documentation items per type of code element.
We can see that w.r.t.~the size of the documentation the three considered APIs are roughly comparable. 
The last row gives the total number of words per API. With an average of 500 words per page, 
the whole API documentation amounts to 2\,800 pages of text.

\label{syntactic-patterns}
Given the sheer size of our API corpus (see Table \ref{descriptive-stats}), it is not possible to read the whole documentation. 
Hence, we decided to first select syntactic patterns that are likely to reveal directives,
and then to analyze a sample of occurrences of those patterns.
For instance, the word ``must'' is a syntactic pattern that indicates important information, and we could analyze 20\% of all paragraphs containing ``must''.

The list of syntactic patterns shown in Table \ref{table-concerns} is the union of: 
\begin{itemize}
\item The patterns mentioned in  RFC \#2119, ``Keywords for use in RFCs to Indicate Requirement Levels''  \cite{Bradner1997}, that is: ``must'', ``should'', ``require*'', ``shall'', ``mandat*'', ``encourage*'',  ``recommend*'', and ``may''. 
\item A list of terms related to inheritance contracts. This enables us to find API directives related to subclassing. This list is based 
upon our work on subclassing directives  \cite{Bruch2010} and encompasses:  ``\code{extend*}'', ``\code{overrid*}'', ``\code{overload*}'', ``\code{overwrit*}'', ``\code{reimplement*}'', ``\code{subclass*}'', ``\code{inherit*}'', and ``\code{super*}''.
\item The patterns described/found in 
 ``Requirements for writing java API specifications'' \cite{Smith2003} (Sun technical report), ``Improving API documentation usability with knowledge pushing'' \cite{Dekel2009a} (conference paper at ICSE), and ``Designing Eclipse APIs'' as well as ``Java API Design'' \cite{Bokowski2006,Bokowski2008} (tutorials by IBM). 
\item Skimming the documentation of 1000 pieces of API documentation randomly selected across the whole corpus.
\end{itemize}
This list contains multiple patterns that can occur in the same sentence or paragraph and may then reveal multiple kinds of directives. Note that skimming the documentation was primarily meant to find syntactic patterns, not kinds of API directives (see Section \ref{manual-analysis}).

\begin{table}
\label{table-concerns}

\begin{tabularx}{\textwidth}{|p{5cm}|X|}
\hline
Category &                                                                             Syntactic Patterns \\
\hline
control flow &                                              call*, invo*, before, after, between, once, prior \\
     generic &                            must, mandat*, require*, shall, should, encourage*, recommend*, may \\
restrictions &       assum*, only, debug*, restrict*, never, condition*, strict*, necessar*, portab*, strong* \\
 performance &                                               performan*, efficien*, fast, quick, better, best \\
 concurrency &                                           concurren*, synchron*, lock*, thread*, simultaneous* \\
 alternative &                                                                 desir*, alternativ*, addition* \\
 subclassing &             extend*, overrid*, overload*, overwrit*, reimplement*, subclass*, super*, inherit* \\
     warning &                                                                   warn*, aware*, error*, note* \\
\hline
\end{tabularx} 
\caption{Concerns and the associated syntactic patterns}

\end{table}

\subsubsection{Finding all Occurrences of a Syntactic Pattern}
To find all occurrences of a syntactic pattern, we indexed
the whole API corpus\footnote{To create and query the index, we used Lucene, see \url{http://lucene.apache.net}}.
Using a web based interface we were then able 
 to query the API corpus in a way similar to Google Search.
For instance, the query ``+content:must +dataset:jface'' returns all occurrences of API documentation in the JFace library that contain ``must''.

\subsubsection{Manual Analysis of API Sentences and Paragraphs and Construction of the Taxonomy}
\label{manual-analysis}
A key part 
is the manual analysis of API sentences and paragraphs that contain occurrences of the syntactic patterns.
For a given occurrence of a syntactic pattern, the analysis consists of reading the whole API comment and to make two decisions. First, whether the context of the occurrence (sentence or paragraph) actually refers to a directive.
Second, whether the taxonomy already contains one or more directive kinds in which the directive can be classified.
The four authors of this paper made this analysis themselves. Each single pattern occurrence was analyzed by one of us, unless the analysis proved to be difficult. 
Indeed, when one of the authors encountered an obscure API comment that is hard to understand/to classify, we discussed it together to make the two aforementioned decisions.
Furthermore, when we came across an API comment "smelling" like a directive, but not fitting in any of the existing kinds of the taxonomy, we discussed the comment together to make sure that it is a directive and -- if so -- to agree on its definition and to find a good name.

\begin{figure}[htpb]
\caption{Searching, reading, and identifying directives is supported with an ad hoc browser-based tool}
\label{fig:javadoxx}
\begin{center}
\includegraphics[scale=0.4]{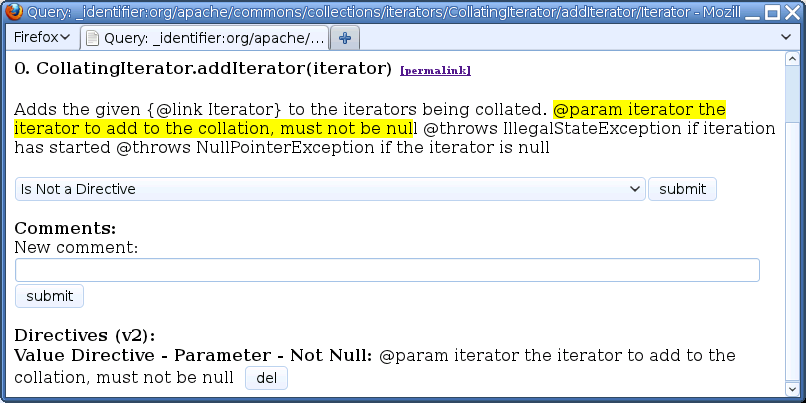}
\end{center}
\end{figure}

To support this analysis, we have designed and implemented a web application that enables the simultaneous, collaborative analysis of API documentation. A screenshot of the application is shown in Figure \ref{fig:javadoxx}. 
The main feature of the application is that it enables to tag pieces of API documentation with a directive kind 
(cf. the select widget and button in Figure \ref{fig:javadoxx}).
To foster discussion, it also supports commenting every piece of API documentation. 
One crucial feature of the tool is that it automatically tracks the links between analyzed occurrences of syntactic patterns and the associated directive kind(s). 

Note that we also permitted tagging directives that were found through serendipity: for instance tagging the next sentence of a pattern occurrence or even tagging directives found with free search over the API documentation database.

\subsubsection{Sampling and Stopping Criterion}
\label{sampling}

For a given dataset and a given syntactic pattern, we use our tool to get the list of all occurrences to be analyzed.
The sampling is then done by automatically shuffling the list and starting with the analysis of the first element. 
Given that the decision whether an occurrence is a directive or not is a binary decision,
the proportion of directives revealed by a pattern is a standard proportion estimation. The stopping criterion is thus the minimum number $MIN$ of occurrences to be analyzed in order to ensure that the estimated proportion is in a certain interval at a certain confidence level. 
The standard statistical formula is \cite[Sec. 3.7]{Singh1996}:
\begin{equation}
MIN = \frac{n_0}{1+\frac{n_0-1}{populationSize}}
\end{equation}
 
$n_0$ depends on the selected confidence level and the desired error margin $e$. $n_{0}$ is calculated using the formulae: $n_0=(Z^2 * 0.25) / e^2$, where Z is a confidence level's z-score. At 95\% confidence level $Z$ is $1.96$. This formula is sometimes also presented as $n_0= 4*(Z^2 * 0.25) / B^2$, where $B$ is the error interval, i.e. $B=2*e$.
When the population size is large enough, MIN tends towards a constant (e.g. 384  for $e=5\%$ at 95\% confidence level); this is known as proportion estimation for large populations.

Our stopping criterion consists of analyzing MIN API elements for $e=10\%$ at 95\% confidence level. For instance, 1448 elements of the JDK's API documentation elements contain at least one occurrence of ``may''. If we analyze $MIN=91$ elements, we are sure that the estimated proportion has a maximum error margin $e$ of 10\% at 95\% confidence level, i.e. $p = x\pm 10\%$ (where $x$ is the measured proportion on 91 items). 

We chose $e=10\%$ 
because it keeps the number of elements to be analyzed at a manageable level (5042 API elements for all 3 datasets and 53 concerns) while still giving us enough confidence in the results. 
Note that for certain keywords, due to a previous version of the case study, we analyzed many more API elements than MIN, as we will see in Section \ref{discussion}.

\subsection{Validity}

\subsubsection{Completeness of the Taxonomy}
\label{taxonomy}

In this section, we discuss whether the taxonomy -- resulting from our exploratory case study -- is complete, i.e. whether we miss an important directive kind.
For this, we list the main threats to validity and discuss the taken counter-measures. 

One possible threat is that the analyzed corpus is too small or is not representative.
However, we think that every API corpus of a certain size and quality will reveal all major directive kinds. Analyzing more APIs would not yield any significant changes in the taxonomy.
Though, defining hard criteria to decide whether an API corpus is large enough and representative is not possible, 
we are still convinced that it is the case for the chosen API corpus for the following reasons.
Our API corpus covers a broad range of different domains (collections, IO, UI, math, SQL, security); and it covers different kinds of API usage such as inheritance or instantiation. Furthermore, the corpus covers three documentation processes employed by three different major software organizations (Sun, Eclipse and Apache) and the documentation has been written by a large number of different authors.

A second threat is that the list of syntactic patterns is too small.
However, our list of syntactic patterns is the result of 4 different processes (see Section \ref{syntactic-patterns}) involving elements from the technical and academic literature, and browsing pieces of API documentation randomly selected across the whole corpus.
Also, the use of patterns may miss some directives, we are confident that it is unlikely that we completely missed a kind of directive.

Third, manual analyses are always error-prone.
To mitigate this threat, we took care that the analysis was made by people with a strong background in (Java) software development. The whole analysis was made by us, the authors of this paper. Three of us hold a doctorate in software engineering, and one of us is a graduate student who is working on her master's thesis. Second, while doing the study, we frequently discussed our intermediate findings and the content of the taxonomy to ensure that everyone has the same understanding of each directive kind.  However, we did not perform any inter-rater reliability measurement.
Even though it is likely that the tagging of some occurrences is at least subjective -- if not even wrong -- the risk of completely missing a directive kind and deriving meaningless results is negligible. 

As a result, we believe that the probability of missing an important directive kind is very low. As for any empirical result, a replication study would be required to increase the confidence in our empirical findings.

\subsubsection{Generalizability of the Taxonomy}
\label{generalizability}
W.r.t.~the generalizability of the taxonomy we believe -- for the same reasons as presented in Section \ref{taxonomy} -- that our API corpus and list of syntactic pattern is large enough and well-designed so that our taxonomy is generalizable to a large extent to all Java APIs, i.e. no major directive kind is missing and no kind is specific to just one API/to our specific API corpus.

Additionally, it is our conviction that the directive kinds emerge from the power and the limitations of the programming language, as well as from the usual design and implementation patterns.
Hence, we assume that the taxonomy should also be valid for languages that are conceptually close to Java, for instance C\#.
But, to validate this claim, further studies and replications are required.

\subsubsection{Reliability and Generalizability of the Abundance}
The process underlying our case-study and the tool to support it enables us to measure the abundance of each directive kind.
However, making strong claims based on the presented measures is not possible. As already outlined in the previous paragraphs, some statements in an API's documentation may or may not be identified as being a directive based on the point-of-view of the analyst or the point-in-time of the analysis. More important -- and as shown by the analysis in Section \ref{abundance} -- the abundance significantly varies across the API corpus. Hence, our measure of  abundance will not hold for all Java APIs.

\section{A Comprehensive Taxonomy of API Directive Kinds}
\label{list-of-directive-kinds}

\begin{figure}[tpb]
\centering\includegraphics[scale=.425]{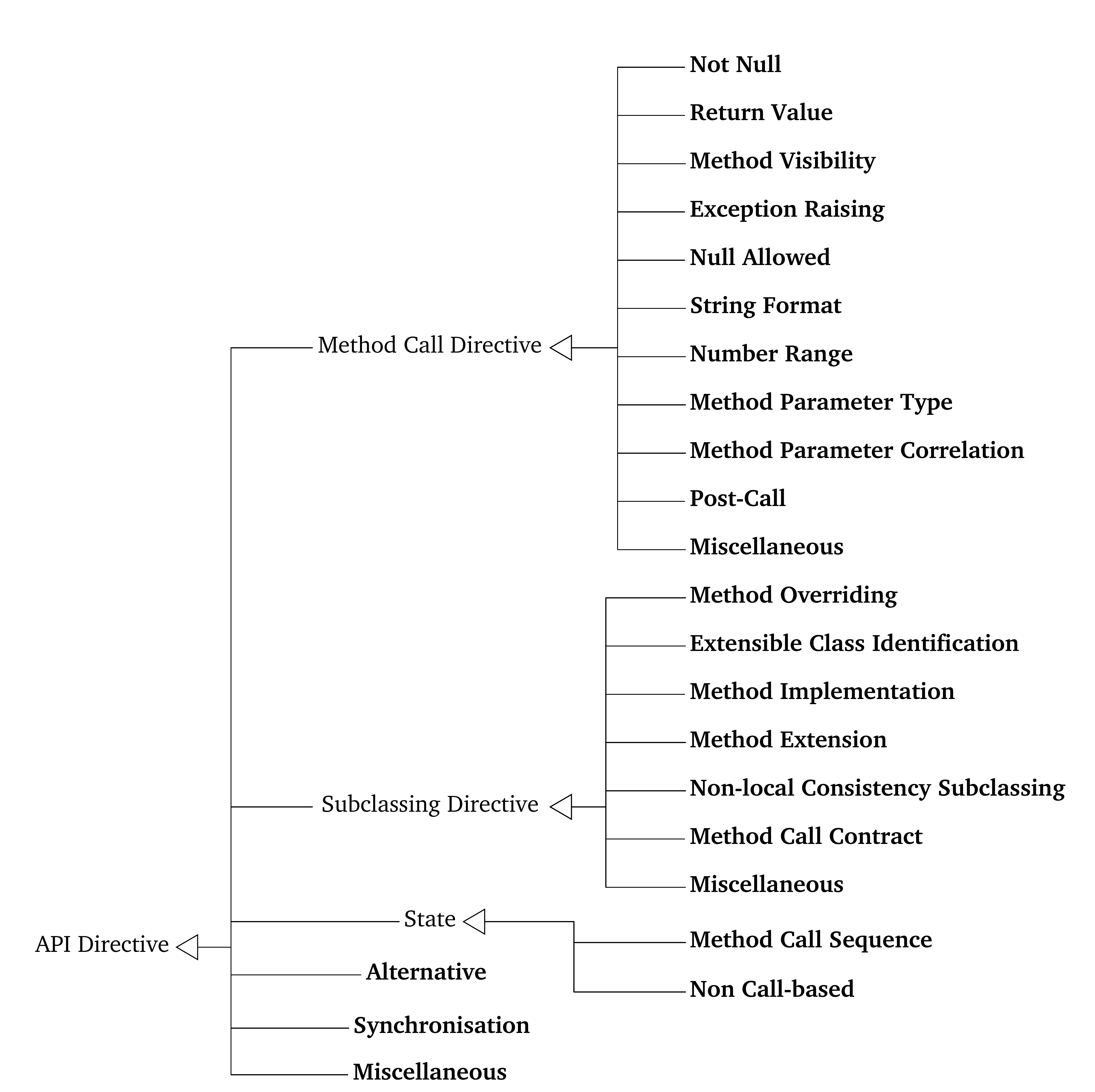}
\caption{Taxonomy of API directives Resulting from of our Empirical Study.}
\label{fig:taxonomy}
\end{figure}

This section discusses the results of the exploratory case study described in the previous section (\ref{methodology}).
We present a comprehensive taxonomy of the kinds of API directives that we found in the analyzed corpuses; the taxonomy is presented in Figure \ref{fig:taxonomy}. Please recall, that we consider as a \textbf{directive}, \emph{a natural-language statement that makes developers aware of constraints and guidelines related to the correct and optimal usage of an API}. Accordingly, a \textbf{directive kind} is \emph{a set of directives that share the same kind of constraints or guidelines}.
For instance, we represent a concrete directive such as: ``@param Document [...] This parameter must not be null'', by a directive kind related to \code{null} values and method parameters, which is called ``Not Null Directive''.

We think that this taxonomy must be described in a systematic manner in order to avoid ambiguity and facilitate understanding and dissemination. To achieve this goal, we  present all directive kinds using the same pattern.
The used pattern is inspired by those used for object-oriented design patterns introduced by \cite{GammaHelmEtAl95} and the pattern used by Cockburn for use-cases \cite{Cockburn2000}.
To some extent, the whole set of directive kinds forms a pattern language to write API documentation \cite{Alexander1977}. 

For each directive kind, we specify the following information:
\begin{description}
 \item[Name] A short name to memorize the directive kind.
 \item[Definition] The explanation of the directive.
 \item[Discussion] The rationales behind the existence of the directive kind and further relevant observations. This also includes -- whenever possible -- the consequences of not being aware of corresponding directives. 
 \item[Prototypical example] A real directive from the corpus that is a prime example for the respective kind of directives.  We chose those examples, from which we think that they are particularly gripping and illustrative.
\end{description}

For some directives, we are able to specify the following additional information: 
\begin{description}
 \item[Use-cases] The different scenarios in which directives of the respective kind are used. Each scenario is described as a use-case.
 
 \item[Good practices] Explanation of good practices to achieve clarity and completeness when describing corresponding directives.
  
 \item[Anti-patterns] Clear anti-patterns that should be avoided when describing directives in the documentation. 

\end{description}

The presentation of directive kinds is ordered by abundance, i.e. the directive kinds that often appear in documentation come first (see section \ref{abundance} about the abundance).

\subsection{Method Call Directive}

\emph{Method call directives} in general express constraints and guidelines when calling a particular library method. Instances of this directive kind, for example, restrict the values input parameters are allowed to take or prescribe how a method's returned value or object has to be used. Given the broad scope of method call directives, we have further refined this directive kind to be able to precisely capture each sub-kind of a method call directive. Each sub-kind is explained in the following sections.

\subsubsection{Not Null Directive}
\label{not-null-directive}
``Not Null'' directives state that a specific method parameter must not be \code{null}.

\begin{description}[leftmargin=0cm]
\item[Discussion] Most object-oriented languages such as Java allow method parameters to be \code{null}.
It is thus possible to assign an application-/context-specific meaning to the \code{null} value. However, there is no default semantics for method parameter values; i.e., users of libraries can neither assume that \code{null} is  allowed nor that it is forbidden. By means of a ``not null directive'' this vagueness is remedied. Not being aware of not null directives often leads to bugs that show up immediately and which are easily identified. 
In many cases, developers directly get a runtime exception (\code{NullPointerException} for Java) if they pass a \code{null} object as method parameter.

\item[Prototypical examples] Using the @param documentation tag: \emph{``@param iterator the iterator to add to the collation, must not be null''}\footnote{in Commons Collections' CollatingIterator.addIterator(iterator)}. Using a tag related to exceptions (@exception, @throws). For instance \emph{``@throws NullPointerException if name is null.''}\footnote{in JDK's RuntimePermission.RuntimePermission(name, actions)}

\item[Anti-pattern] Each library should adopt one strategy where and how the constraint is specified that a certain method parameter must not be null. For example, a library should not sometimes specify the directive  using an @param tag and sometimes using an @throws tag or specify it as part of the general documentation. 

\end{description}

\subsubsection{Return Value Directive}

A ``Return Value'' directive states a required property on the object or value that is returned by a method implementation.
A ``Return Value'' directive makes explicit what client code can assume.

\begin{description}[leftmargin=0cm]

\item[Discussion] It is important for client code to know which assumption it can make on returned objects and values. This significantly reduces the amount of defensive code (such as a not-null check).
When the directive is attached to an abstract or interface method, a return value directive is a contract that implementors must satisfy. When it is attached to a concrete method with an implementation, it documents which property the return value has. In the former case, this directive kind is also a subclassing directive.

Some return value directives can become quite complex: \emph{``The returned information control creator must create information controls  that implement IInformationControlExtension3 and IInformationControlExtension2,  and IInformationControlExtension2.setInput(Object) accepts all inputs that are  also supported by this information control.''\footnote{in JFace's IInformationControlExtension5.getInformationPresenterControlCreator()}}

\item[Prototypical examples] 
  To specify the required precision: \emph{``The computed result must be within 1 ulp of the exact result.''\footnote{in JDK's Math.cos(a)}};
  To specify that \code{null} is a possible return value: \emph{``Returns the receiver's parent, which must be a Shell  or  null.''\footnote{in JFace's Dialog.getParent()}};
  To specify ordering: \emph{``@return an array of integer arrays of 2 elements. Ranges must be in ascending order and must not overlap..''\footnote{in JDK's JobAttributes.getPageRanges()}};
  To specify a relationship between the return value and a parameter: \emph{``An implementation of this method must either return a class  with the given name or throw an exception.''\footnote{in JDK's RMIClassLoaderSpi.loadClass(codebase, name, defaultLoader)}};
  To specify the internal state of the returned object: \emph{``The returned control's layout data must be an instance of GridData.''\footnote{in JFace's PopupDialog.createDialogArea(parent)}}.

\end{description}

\subsubsection{Method Call Visibility Directive}

A ``Method Call Visibility Directive'' is a directive that defines the visibility of a method.
Contrary to state directives (see Section \ref{state-directives}), they do not put restrictions on the state of the receiver of the method call, but on the calling context.

\begin{description}[leftmargin=0cm]

\item[Discussion] Given standard language features such as packages and visibility modifiers, it is not always possible to precisely define the allowed/supported calling contexts.
The visibility is sometimes a permission (e.g. \emph{``you may call this method''}) and sometimes a restriction (e.g. \emph{``you should not call this method''}).
The visibility is sometimes indirect: for instance, in a class-level documentation item, \emph{``This class may be instantiated''} actually means that the constructor or the factory method is visible.

\item[Prototypical examples] 
  \emph{``It should not be called from application code.''\footnote{in JDK's Logger.setParent(parent)}},
  \emph{``Constructor only used in deserialization, do not use otherwise.''\footnote{in Commons Collections' AbstractOrderedMapDecorator.AbstractOrderedMapDecorator()}},
  \emph{``This method is internal to the framework; subclassers should not call this method.''\footnote{in JFace's ListEditor.getShell()}}.

\end{description}

\subsubsection{Exception Raising Directive}

An ``Exception Raising Directive'' states a requirement on the exceptions thrown by a method implementation. The requirement may concern:
1) the type of an exception that may be thrown;
2) the situations in which an exception has to be or is thrown;
3) an exception that must never be thrown.

\begin{description}[leftmargin=0cm]
\item[Discussion] 
Java supports so-called checked exceptions and unchecked exceptions. In the former case, exception raising requirements of subclasses of \code{Exception} can be expressed using \code{throws} declarations in method signatures. However, requirements on others kinds of exceptions have to be expressed as exception raising directives.
Although developers are necessarily aware of checked exceptions by the programming language, it is still necessary to document them to describe when and why they are thrown.
Exception directives are often documented with \texttt{@exception} or \texttt{@throws} Javadoc tags.

\item[Prototypical examples] 
  \emph{``@exception IllegalArgumentException if groupID is null.''\footnote{in JDK's ActivationDesc.ActivationDesc(groupID, className, location, data, restart) }},
  \emph{``This constructor also must throw an IllegalArgumentException if it does not understand the parameters input.''\footnote{in JDK's PolicySpi}},
  \emph{``DOM applications must not raise exceptions in a filter.''\footnote{in JDK's LSParserFilter}}.

\item[Anti-pattern] Stating an exception raising directive that makes several implementation strategies possible.
For instance, \emph{``@throws IllegalStateException implementations may, but are not required to, throw this exception if the entry has been removed from the backing map.''\footnote{in JDK's Map.Entry.getKey()}}. In this case, a developer who calls the interface method has no guarantee how the method will behave.

\end{description}

\subsubsection{Null Allowed Directive}
A ``Null Allowed'' directive specifies that a method parameter is allowed to be \code{null} and explains the specific semantics of the \code{null} value for the respective parameter.

\begin{description}[leftmargin=0cm]

\item[Discussion] This is the dual directive of the ``Not Null'' directive (see Section \ref{not-null-directive}). 
In contrast to the ``Not Null'' directive, nothing bad will happen if a developer misses this directive.  
The developer will simply miss some functionality provided by the method.

\item[Prototypical examples] \emph{``null means that the catalog name should not be used to narrow the search''\footnote{In JDK's DatabaseMetaData.getColumnPrivileges(catalog, schema, table, columnNamePattern)}}, \emph{``If this parameter is null, then only a specified keyStroke will invoke content proposal''\footnote{In JFace's ContentProposalAdapter.setAutoActivationCharacters(autoActivationCharacters)}},

\item[Anti-pattern] Stating that \code{null} is allowed (e.g. \emph{``may be null''}), but not specifying the semantics of passing \code{null}.
\end{description}

\subsubsection{String Format Directive}
A ``String Format'' directive prescribes the allowed format of a string that is passed in as a method parameter.

\begin{description}[leftmargin=0cm]

\item[Discussion] In general, it is a best practice to use structured objects rather than strings as method parameters (cf. \cite[page 50]{Bloch2008}, ``Avoid strings where other types are more appropriate''). However, in some cases using a plain \code{String} object makes the client code much more concise and readable. The two prime examples, where strings are usually used as input parameters are regular expressions and the \code{printf} method and its variants. 
As in case of the ``Number Range'' directive, we did find instances where the use of an enumeration where each enum-value is associated with a string, would be an improvement. We further found instances of ``String Format'' directives where there is an obviously more appropriate type (such as \code{File} for denoting files\footnote{Runtime.load(filename): \emph{``The filename argument must be a complete path name, (for example Runtime.getRuntime().load("/home/avh/lib/libX11.so");)''}}).

\item[Prototypical examples]  \emph{``The name must conform to RFC 2965.''\footnote{in JDK's HttpCookie.HttpCookie(name, value)}},
\emph{``@param name Permission name. Must be either "monitor" or "control".''\footnote{in JDK's  ManagementPermission.ManagementPermission(name, actions). Although this may seem to be a bad design, we include it here since examples like this do appear in the APIs we studied.}}

\end{description}

\subsubsection{Number Range Directive}

A ``Number Range'' directive states that only certain values of a number's domain are allowed to be used.
The directive is applicable to method parameters that represent numbers --- e.g. byte, int, long, short, float, double and the corresponding wrapper classes (Integer, BigInteger, etc.)

\begin{description}[leftmargin=0cm]
\item[Discussion] In some cases a method parameter is only allowed to take a value in a specific range. For example, to specify the id of a port on which to wait for incoming network connections (e.g., the port id must be in the range [0,...,65535]). Given that it is in Java (as well as in other mainstream programming languages) not possible to define a (value) type that only allows values in a specific range, a value type is chosen that can accommodate the corresponding values and an additional API directive is specified. The directive then specifies the restriction on the allowed values.
We did found a number of instances of number range directives, where 
the range is a strictly finite set of discrete values and where each value has a very specific meaning. In such cases it would be advisable to use an Enumeration (e.g., using Java's ``enum'' feature) to improve the comprehensibility and readability of the code.
E.g. \emph{``@param autoGeneratedKeys one of the following constants: Statement.RETURN\_GENERATED\_KEYS or Statement.NO\_GENERATED\_KEYS''\footnote{in JDK's Statement.execute(sql, autoGeneratedKeys)}}.

\item[Prototypical examples] 
\emph{``The port must be between 0 and 65535, inclusive.''\footnote{in JDK's  ServerSocket.ServerSocket(port, backlog, bindAddr)}},
\emph{``@param scale the desired number of digits to the right of the decimal point. It must be greater than or equal to zero.''\footnote{in JDK's  CallableStatement.registerOutParameter(parameterIndex, sqlType, scale)}}

\end{description}

\subsubsection{Method Parameter Type Directive}
A ``Method Parameter Type Directive'' restricts the allowed type of method parameters.

\begin{description}[leftmargin=0cm]
\item[Discussion] Type systems are generally not able to express all meaningful/desired type constraints. 
We did found instances of this directive where it was used to state that the method violates the contract of the superclass.
For instance, \code{java.sql.Timestamp} which inherits from \code{java.util.Date} states that it violates the contract of the method \code{compareTo} which accepts a \code{Date} object as parameter. The \code{compareTo} method of the subclass states: \emph{``Compares this Timestamp object to the given Date, which must be a Timestamp object.''}

\item[Prototypical examples] 
  \emph{``@param obj must be a Number or a Date.''\footnote{in JDK's DateFormat.format(obj, toAppendTo, fieldPosition)}},
  \emph{``The CertPath specified must be of a type that is supported by the validation algorithm, otherwise an InvalidAlgorithmParameterException will be thrown.''\footnote{in JDK's CertPathValidator.validate(certPath, params)}},
  \emph{``@param obj [typed as Object] the object to be serialized (must be serializable) [i.e. must implement the interface Serializable]''\footnote{in JDK's MarshalledObject.MarshalledObject(obj)}}.

\end{description}

\subsubsection{Method Parameter Correlation Directive}

A ``Method Parameter Correlation Directive'' describes inter-dependencies involving two or more parameters of a method.

\begin{description}[leftmargin=0cm]
\item[Discussion] There is no language construct to express those dependencies. However, it sometimes reveals a bad design that could be fixed by means of assigning more appropriate parameter types and method overloading.

\item[Prototypical exampless] \emph{``If the given key is of type java.security.PrivateKey,  it must be accompanied by a certificate chain certifying the  corresponding public key.''\footnote{in JDK's KeyStore.setKeyEntry(alias, key, password, chain)}},
\emph{``The reader must contain  the number  of characters specified by length otherwise a SQLException will be  generated when the CallableStatement is executed.''\footnote{in JDK's CallableStatement.setNClob(parameterName, reader, length)}}.

\end{description}

\subsubsection{Post-Call Directive}

A ``Post-Call'' directive prescribes what immediately needs to be done with the returned object.
 It is a special case of a ``Method-call Sequence'' directive (see Section \ref{method-call-sequence-directive}) 
 which does not involve multiple method calls and which is generally related to finishing the setup/initialization of the returned object.

\begin{description}[leftmargin=0cm]
\item[Discussion] It is not always possible to encapsulate all behavior in a single method, especially when the behavior depends on client code. In such cases, a ``Post-Call'' directive states what remains to be done.

\item[Prototypical examples]
  \emph{``The returned parameter object must be initialized via a call to init, using an appropriate parameter specification or parameter encoding.''\footnote{in AlgorithmParameters.getInstance(algorithm)}},
  \emph{``The return value should always be checked to make sure that the rename operation was successful.''\footnote{in JDK's File.renameTo(dest)}}

\end{description}

\subsubsection{Miscellaneous Method Call Directives} 

This section presents method call directives that can not be classified using the aforementioned directive kinds and which were too rare and to fuzzy to warrant a directive kind on their own (less than 20 items on the whole corpus).

\begin{description}[leftmargin=0cm]

\item[Complex directive involving the type and the history of the parameter] 
\emph{``@param obj the server object for which a stub is required. Must either be a subclass  of PortableRemoteObject or have been previously the target of a call to exportObject''\footnote{in JDK's PortableRemoteObject.toStub(obj)}}

\item[Complex directive involving the application life-cycle and concurrency issues] 
\emph{``Shutdown hooks run at a delicate time in the life cycle of a virtual machine and should therefore be coded defensively. They should, in particular, be written to be thread-safe and to avoid deadlocks insofar as possible. They should also not rely blindly upon services that may have registered their own shutdown hooks and therefore may themselves be in the process of shutting down.''\footnote{in JDK's Runtime.addShutdownHook(hook)}}

\end{description}

\subsection{Subclassing Directive}
\label{subclassing-directives}

A ``Subclassing'' directive states a requirement that has to be satisfied when subclassing a library class; such directives are, e.g., related to
how and when to extend a library class, implementing a library interface, overriding a library method, implementing an abstract library method (or a method from a library interface).
This section deepens the empirical findings of our previous work  \cite{Bruch2010}, which introduced the concept of ``subclassing directives''.

\subsubsection{Method Overriding Directive}
\label{method-overriding-directives}
A ``Method Overriding'' directive specifies whether a method is intended or expected to be overridden by client subclasses. Here, overriding is used in its broadest sense: implementing an abstract method, replacing the default implementation, or extending the default implementation by calling \code{super}.
It can be constructive (e.g. ``may'' or ``should'') or restrictive (e.g. ``should not'').

\begin{description}[leftmargin=0cm]

\item[Discussion] Method overriding directives tell client developers what methods are intended or expected to be overridden, but using standard (Java) language features, it is not possible to express general method overriding directives. In particular, protected methods can always be overridden by library subclasses and client subclasses. 
However, library designers often need to enable subclasses to override a specific method,  but simultaneously want to maximize the encapsulation and, hence, do not want client code to override a specific method.

\item[Prototypical examples] 
  \emph{``The subclass of ActivationGroup must override this method and unexport the object.''\footnote{in JDK's ActivationGroup.inactiveObject(id)}},
  \emph{``Subclasses may override any of the following methods: isLabelProperty, getImage, getText, dispose''\footnote{in JFace's LabelProvider}},
  \emph{``Subclasses can override this to match differently.''\footnote{in Apache Commons Collections' AbstractLinkedList.isEqualValue(value1, value2)}}.

\item[Anti-pattern]
Using terminology related to the overriding of methods (override, extend, re-implement, overwrite and implement), without a strict agreement on the precise semantics. 
Based on our experience, it seems that some developers understand ``override'' as discarding the library implementation (i.e. not calling super) and other understand ``extend'' as a requirement to call the library implementation using a super call. However, both uses do have counter-examples and this can lead to confusion.

\end{description}

\subsubsection{Extensible Class Identification Directive}
\label{extensible-class-identification}

An ``Extensible Class Identification'' directive indicates to what extent classes and interfaces of a library can be subclassed by developers. 
It can be constructive (e.g. ``may'' or ``should'') or restrictive (e.g. ``should not'').

\begin{description}[leftmargin=0cm]
\item[Discussion] In object-oriented libraries, there is generally a mix of classes to be instantiated and classes to  be subclassed. Even if the API only contains public classes (i.e.~which are allowed to be used by clients), developers need to know whether the main usage of the API classes is instantiation or inheritance.

\item[Prototypical examples]
  \emph{``If the  programmer desires thread-local variables to have an initial  value other than null, ThreadLocal must be subclassed, and this method overridden.''\footnote{in JDK's ThreadLocal.initialValue()}},
  \emph{``For providing the label's styles, create a subclass and overwrite update(ViewerCell) to set all information needed to render a element.''\footnote{in JFace's StyledCellLabelProvider}}

\item[Anti-pattern]
Designating a class as extensible, without stating the condition or the scenario in which subclassing happens. E.g.\emph{``This class is intended to be subclassed by implementors.''} or \emph{``An abstract implementation of a linked list which provides numerous points for subclasses to override.''\footnote{in Apache Commons Collections' AbstractLinkedList}}.

\end{description}

\subsubsection{Method Implementation Directive}

A ``Method Implementation'' directive states a requirement or contract on the implementation of a given method that is related neither to the returned value nor to raising exceptions.

\begin{description}[leftmargin=0cm]

\item[Discussion] This directive expresses general contracts for method implementors; it generally applies to abstract methods, interface methods and methods that are intended to be overridden.

\item[Prototypical examples] 
  \emph{``If the scale is reduced by the operation, the unscaled value must be divided (rather than multiplied), and the value may be changed''\footnote{in JDK's BigDecimal.setScale(newScale, roundingMode)}},
  \emph{``This method must not modify the parent's layout.''\footnote{in JFace's Dialog.createDialogArea(parent}},
  \emph{``According to the Collection.remove(Object) method, this method should only remove the first occurrence of the given object, not all occurrences.''\footnote{in Apache Commons Collections' Bag.remove(object)}}.

\item[Anti-pattern] Writing the directive such that it can be confused with an implementation comment, e.g. "Returns an item taken from the internal list" instead of "Must return an item that is taken from the internal list". This is particularly important for methods that already have a body, since there is no way to differentiate between implementation comments and contracts.

\end{description}

\subsubsection{Method Extension Directive}
\label{extension-directive}

A ``Method Extension'' directive states whether an application-level method -- overriding a library method -- has to execute the default implementation by calling super.

\begin{description}[leftmargin=0cm]
\item[Discussion] In practice, it is not always possible or desirable to design extension points by means of abstract methods. However, the code of an overridable method can contain important functionality with respect to the control-flow or the state of the library. Method extension directives document this fact. 

\item[Prototypical examples] 
  \emph{``Subclasses may override but must call super.doFillIntoGrid.''\footnote{in JFace's StringFieldEditor.doFillIntoGrid(parent, numColumns)}},
  \emph{``Code overriding this method should call super.removeNotify as the first line of the overriding method.''\footnote{in JDK's Component.removeNotify()}},
  \emph{``Subclasses which override this method should make sure they call super''\footnote{in Apache Commons Collections' AbstractLinkedList.init()}}.

\end{description}

\subsubsection{Non-local Consistency Subclassing Directive}

A ``Non-local Consistency Subclassing''  directive states that clients have to satisfy consistency requirements that span multiple overridable methods. For instance, a consistency subclassing directive may state that two methods have to be overridden together.

\begin{description}[leftmargin=0cm]

\item[Discussion] For some reason (e.g. readability, modularity, reusability), certain functionality is  implemented using several methods or classes, but there is a general contract -- e.g.~an invariant spanning those elements --  that needs to be satisfied. 

\item[Use-cases] 
  When two methods have to be overridden together, e.g.~\emph{``The visibility of the details button is controlled by shouldShowDetailsButton(), which should also be overridden together with this method.''\footnote{in JFace's ErrorDialog.createDropDownList(parent)}}~or \emph{``Note that it is generally necessary to override the hashCode method whenever this method is overridden, so as to maintain the general contract for the hashCode method, which states that equal objects must have equal hash codes.''\footnote{in JDK's Object.equals(obj)}}. 
  When two classes have to be overridden together, e.g.~\emph{``Concrete subclasses of ColumnViewer should implement a matching concrete subclass of ViewerColumn.''\footnote{in JFace's ViewerColumn}}
  When a method has to be overridden together with a client implementation of a library interface, e.g.~\emph{``Clients may implement this interface and override TextViewer.create\-Do\-cu\-mentAdapter if they want to intercept the communication between the viewer's text widget and the viewer's document.''\footnote{in JFace's IDocumentAdapter}}
  When the returned value of a method (here \code{format}) must be in the range of valid input values of another method (here \code{parseObject}), e.g.~\emph{``Generally, a format's parseObject method must be able to parse any string formatted by its format method.''\footnote{in JDK's Format}}.

\end{description}

\subsubsection{Call Contract Subclassing Directive}

A ``Call Contract Subclassing Directive'' directive states whether an application-level method -- overriding a library method -- has to execute certain library methods other than just calling the \code{super} method.

\begin{description}[leftmargin=0cm]

\item[Discussion] As for ``Method Extension'' directives, the control-flow or the state of a library sometimes has to be maintained through client-level calls to library methods. Method call directives document these requirements.

\item[Prototypical examples] 
  \emph{``Subclasses must call this method at the end of their constructor(s)''\footnote{in JFace's AbstractInformationControl.create()}}, \emph{``If this method is overridden, the method that overrides it should additionally check to see if the calling thread has the RuntimePermission("modifyThread") permission, and if so, return silently.''\footnote{in JDK's SecurityManager.checkAccess(t)}}.

\item[Anti-pattern]
Being vague either on the method to be called or on the context in which the call must happen, for instance \emph{``The message, image and title should be updated by the subclass''\footnote{in JFace's ErrorDialog}} puts the responsibility of figuring out the corresponding methods on the developer who uses the dialog. A second example is:  
\emph{``All subclasses must call this method when their control is first established.''}. In this case the questions is: ``Where and how is the control generally established?''

\end{description}

\subsubsection{Miscellaneous Subclassing Directives}
This section presents other kinds of subclassing directives that can not be classified in the aforementioned directive kinds and which are rare (less than 20 items on the whole corpus). 

\begin{description}[leftmargin=0cm]

\item[Documentation Directive] Such directives state that the implementation must document certain pieces of information (which can themselves be directives). For instance, \emph{``An implementation is required to clearly document the semantics and guarantees provided by each of the waiting methods''\footnote{in JDK's Condition}} or \emph{``Each class that implements PrivilegedExceptionAction should document what (if anything) this value represents.''\footnote{in JDK's PrivilegedExceptionAction.run()}}

\item[Directive on the required constructor of a class] 
\emph{``The class [specified by \code{java.system.class.loader}] is loaded using the default system class loader  and must define a public constructor that takes a single parameter of  type ClassLoader which is used as the delegation parent.\footnote{in JDK's ClassLoader.getSystemClassLoader()}}

\end{description}

\subsection{State Directives}
\label{state-directives}

A ``State Directive'' is a requirement on the internal state of receivers of a given method call.
Most state directives that we found are expressed as method call sequences; i.e.~first method \code{A} has to be called before method \code{B} can be called/has to be called (see Section \ref{method-call-sequence-directive}). But, some are more complex and have state related constraints that are beyond method call sequences.

\subsubsection{Method Call Sequence Directive}
\label{method-call-sequence-directive}

A ``Method Call Sequence'' directive specifies an object usage protocol. In particular, such directives specify the order of method calls.

\begin{description}[leftmargin=0cm]

\item[Discussion] 
In many cases, a method can only be called when an object is in a certain state. Though approaches were proposed to formally specify method object protocols as part of the source code or documentation \cite{Bierhoff2009,Bierhoff2011}, no support exists in standard programming languages. 

\item[Prototypical example] \emph{``The getKeyStore method must be invoked before this method may be called.''\footnote{in JDK's KeyStore.Builder.getProtectionParameter(alias)}}

\item[Use-cases] 
  Before a method can be used for the first time, its initialization has to be finished, e.g., \emph{``You must add at least one Comparator before calling the compare(Object,Object) method''\footnote{in Apache Commons Collections' ComparatorChain.ComparatorChain()}}.
  Certain method call sequence are conditional, e.g.~\emph{``Should be called whenever needsInput() returns true indicating that more input data is required.''\footnote{in JDK's Inflater.setInput(b)}}.
  To restrict the number of times a method is called, e.g.~\emph{``This method may only be called once; the call must occur before JFaceResources.getFontRegistry is invoked.''\footnote{in JFace's JFaceResources.setFontRegistry(registry)}}.

\item[Anti-pattern] Not precisely stating the methods that are involved in the sequence, for instance \emph{``A group must first be registered with the ActivationSystem before it can be created via this method''\footnote{in JDK's ActivationGroup.createGroup(id, desc, incarnation)}} or \emph{``For client sockets, setReceiveBufferSize() must be called before connecting the socket to its remote peer.''\footnote{in JDK's Socket.setReceiveBufferSize(size)}}.

\end{description}

\subsubsection{Non Call-based State Directive}
\label{complex-state-directives}

A ``Non Call-based  State'' directive states a requirement on the application state that is not expressible as a method call sequence. If the application is in the specified state certain method calls are allowed to be made.

\begin{description}[leftmargin=0cm]

\item[Discussion] Same arguments as in Section \ref{method-call-sequence-directive}.

\item[Use-cases] 
  To express that certain resources associated with the object have to be available at the point in time of calling the method, for instance \emph{``[This object] is valid only during the paint and must not be disposed''\footnote{in SWT's Event.gc}}.
  To refer to the state of external resources, e.g.~the file system \emph{``If this pathname denotes a directory, then the directory must be empty in order to be deleted.''\footnote{in JDK's File.delete()}}.
  To express a requirement on the ownership of an object (that can not be expressed in Java), for instance \emph{``@param caller the principal invoking this method. It must be an owner of this ACL.''\footnote{in JDK's Acl.removeEntry(caller, entry)}} or \emph{``The current thread must own this object's monitor.''\footnote{in JDK's Object.wait(timeout)}}

\end{description}

\subsection{Alternative Directive}

An ``Alternative'' directive states that there are alternative implementations of a given API element.

\begin{description}[leftmargin=0cm]

\item[Discussion] Often, in the same API; multiple ways exist to achieve the same goal, each of them making different trade-offs.

\item[Use-cases] 
  To specify the method or class that replaces a deprecated method or class. (The widely used @deprecated tag is on its own not sufficient, e.g. \emph{``@deprecated since 3.1 use {@link org.eclipse.jface.text.DefaultIndentLineAutoEditStrategy} instead''\footnote{in JFace's DefaultAutoIndentStrategy}}).
  To indicate a more efficient implementation, e.g.~\emph{``The StringBuilder class should generally be used in preference to this one, as it supports all of the same operations but it is faster, as it performs no synchronization.''\footnote{in JDK's StringBuffer}}

\item[Anti-pattern] Stating an alternative directive with no rationales or conditions, e.g. \emph{``Subclassers are advised to override inputChanged rather than this method''\footnote{in JFace's ContentViewer.setInput(input)}}. Such directives raise more questions than they answer.

\item[Good practice] Alternative directives often consist of only two alternatives: the ``better'' and the ``unrecommended'' one. A good practice is then to state the alternative directive in the API documentation at least in the ``unrecommended'' element to minimize the risk of missing the directive.

\end{description}

\subsection{Synchronization Directive}

A ``Synchronization'' directive states some information regarding the impact of concurrency 
on an API element.

\begin{description}[leftmargin=0cm]

\item[Discussion] 
Synchronization directives state what needs to be considered or needs to be done when using a specific API element in a concurrent program. 
Often, this kind of directive is expressed in a passive manner, such as \emph{``If multiple threads access a hash map concurrently, and at least one of the threads modifies the map structurally, it must be synchronized externally.''\footnote{in JDK's HashMap}}.

\item[Use-cases] 
  To warn that an API element is not ready for use in concurrent environments, i.e.~there is a risk of race conditions; for instance \emph{``Note that FixedSizeMap is not synchronized and is not thread-safe. If you wish to use this map from multiple threads concurrently, you must use appropriate synchronization.''\footnote{in Apache Commons Collections' FixedSizeMap}} 
  To give concrete guidelines on how to solve library related concurrency problems (e.g.~by acquiring a lock, checking the status of a lock, or creating a synchronized block); for instance, \emph{``Implementations must synchronize on the hierarchy lock.''\footnote{in JDK's BeanContextSupport.addAll(c)}}
  To state that the implementors are responsible for correctly handling concurrency, e.g.~\emph{``The public methods of all CertStoreSpi objects must be thread-safe.''\footnote{in JDK's CertStoreSpi}}. In this case, it is a kind of subclassing directive as exposed in Section \ref{subclassing-directives}.

\end{description}

\subsection{Miscellaneous Directive}

Interestingly, certain directives present in the API documentation are not related to the API usage itself, but to the software environment in general. Here are two kinds of such directives, which have been encountered several times.

\begin{description}[leftmargin=0cm]

\item [Directive for implementors of Java compilers] \emph{``Compilers must ignore any warning names they do not recognize.''\footnote{in JDK's SuppressWarnings.value()}}

\item[Directives related to elements external to the API] \emph{``The driver's behavior must be consistent with a  particular DBMS, either always continuing to process commands or never  continuing to process commands.''\footnote{in JDK's Statement.executeBatch()}}. Note that this consistency directive involves two different kinds of collaborators: the Java library itself, and the external DBMS.

\end{description}

\section{Findings}
\label{discussion}

In this section, we discuss the other findings from our exploratory case study. We provide insights on the suitability of syntactic patterns for finding directives (\ref{precision}), and on the abundance of each directive kind (\ref{abundance}). We also further compare our taxonomy against Dekel's one (\ref{comparison-dekel}) and provide a first set of guidelines on how to write API directives based on our experience. Note that we already discussed the  completeness and the generalizability of our results in Section \ref{methodology}.

\begin{table}
\begin{tabularx}{\textwidth}{|X|X|X|X|}

 \hline
           Concern &               Java &              JFace & commons.collections \\
\hline
              must & 78.6$\pm$0.3\% (740) & 94.4$\pm$5.4\% (179) & 98.2$\pm$0.3\% (330) \\
           mandat* & 42.9$\pm$0.0\% (7) &              - (0) & 25.0$\pm$0.0\% (4) \\
          require* & 70.3$\pm$3.4\% (232) & 58.3$\pm$6.2\% (72) & 29.6$\pm$6.9\% (27) \\
            should & 67.5$\pm$0.3\% (750) & 83.9$\pm$5.7\% (205) & 76.4$\pm$5.6\% (55) \\
        encourage* & 76.2$\pm$8.7\% (21) & 100.0$\pm$0.0\% (6) & 100.0$\pm$0.0\% (4) \\
        recommend* & 88.4$\pm$8.9\% (43) & 95.5$\pm$9.2\% (22) &  0.0$\pm$0.0\% (1) \\
               may & 60.5$\pm$5.5\% (258) & 93.7$\pm$3.3\% (351) & 57.1$\pm$7.5\% (84) \\
           extend* & 22.6$\pm$8.7\% (62) & 68.7$\pm$0.0\% (163) & 16.7$\pm$4.1\% (24) \\
          overrid* & 90.8$\pm$6.3\% (130) & 85.1$\pm$0.7\% (249) & 94.4$\pm$0.0\% (54) \\
         overload* & 57.1$\pm$9.6\% (14) & 100.0$\pm$0.0\% (2) &              - (0) \\
         overwrit* & 10.0$\pm$9.8\% (10) & 75.0$\pm$0.0\% (20) &  0.0$\pm$0.0\% (2) \\
      reimplement* &  0.0$\pm$0.0\% (2) & 97.4$\pm$0.0\% (39) &              - (0) \\
         subclass* & 77.0$\pm$6.6\% (139) & 94.5$\pm$0.6\% (452) & 81.7$\pm$5.4\% (60) \\
            super* & 79.0$\pm$8.4\% (62) & 69.6$\pm$4.6\% (46) & 20.0$\pm$8.9\% (15) \\
          inherit* & 28.9$\pm$9.8\% (38) &  0.0$\pm$0.0\% (1) &              - (0) \\
             note* & 66.3$\pm$9.1\% (101) & 67.6$\pm$9.3\% (74) & 77.1$\pm$6.4\% (48) \\
         efficien* & 72.9$\pm$9.7\% (48) & 55.0$\pm$9.1\% (20) & 50.0$\pm$0.0\% (12) \\
             reus* & 21.4$\pm$6.7\% (28) & 54.5$\pm$8.9\% (11) &  0.0$\pm$0.0\% (7) \\
            desir* & 49.2$\pm$9.7\% (59) & 51.7$\pm$10.0\% (29) & 100.0$\pm$0.0\% (2) \\
       alternativ* & 45.5$\pm$9.2\% (33) & 63.2$\pm$9.6\% (19) & 85.7$\pm$0.0\% (7) \\
         addition* & 46.0$\pm$10.0\% (63) & 31.9$\pm$9.5\% (47) & 7.3$\pm$2.4\% (41) \\
             warn* & 26.9$\pm$9.9\% (26) & 25.0$\pm$6.8\% (20) & 81.8$\pm$0.0\% (11) \\
            aware* & 66.7$\pm$0.0\% (9) & 27.3$\pm$0.0\% (11) &              - (0) \\
            error* & 89.3$\pm$9.3\% (103) & 38.5$\pm$10.0\% (52) & 50.0$\pm$0.0\% (16) \\
        concurren* & 79.0$\pm$9.1\% (81) &  0.0$\pm$0.0\% (1) & 87.5$\pm$0.0\% (32) \\
         synchron* & 87.2$\pm$8.6\% (78) & 11.1$\pm$0.0\% (9) & 74.6$\pm$3.7\% (71) \\
             lock* & 63.2$\pm$9.8\% (68) & 16.7$\pm$0.0\% (6) & 75.0$\pm$0.0\% (12) \\
           thread* & 65.9$\pm$6.2\% (179) & 78.4$\pm$8.5\% (37) & 70.5$\pm$5.5\% (44) \\
              fast & 73.9$\pm$7.1\% (23) & 100.0$\pm$0.0\% (2) & 20.0$\pm$0.0\% (5) \\
             quick &  0.0$\pm$0.0\% (2) & 22.6$\pm$9.9\% (31) &              - (0) \\
             call* & 75.2$\pm$4.8\% (343) & 73.5$\pm$4.6\% (260) & 56.3$\pm$9.5\% (71) \\
             invo* & 68.0$\pm$8.9\% (103) & 41.7$\pm$8.5\% (48) & 62.9$\pm$9.2\% (35) \\
        performan* & 45.7$\pm$8.8\% (46) & 75.0$\pm$0.0\% (4) & 53.8$\pm$0.0\% (13) \\
         restrict* & 66.0$\pm$9.8\% (47) & 50.0$\pm$0.0\% (4) & 50.0$\pm$0.0\% (2) \\
             never & 60.0$\pm$9.5\% (60) & 78.3$\pm$9.9\% (46) & 76.9$\pm$0.0\% (13) \\
        condition* & 61.3$\pm$9.9\% (62) &  0.0$\pm$0.0\% (6) &  0.0$\pm$0.0\% (1) \\
           strict* & 50.0$\pm$10.0\% (34) &              - (0) &  0.0$\pm$0.0\% (1) \\
         necessar* & 58.6$\pm$9.9\% (70) & 30.8$\pm$9.5\% (26) & 13.0$\pm$4.3\% (23) \\
           portab* & 92.9$\pm$9.6\% (14) &              - (0) &              - (0) \\
           strong* & 42.9$\pm$10.0\% (28) &              - (0) &              - (0) \\
            assum* & 42.0$\pm$9.6\% (50) & 52.9$\pm$10.0\% (34) & 29.4$\pm$0.0\% (17) \\
              only & 77.5$\pm$7.7\% (138) & 68.4$\pm$9.6\% (76) & 40.7$\pm$2.8\% (91) \\
            debug* & 11.5$\pm$9.0\% (26) & 41.2$\pm$0.0\% (17) &  0.0$\pm$0.0\% (4) \\
            before & 74.2$\pm$9.0\% (89) & 52.5$\pm$9.5\% (59) & 35.7$\pm$3.3\% (42) \\
             after & 60.7$\pm$9.5\% (84) & 57.4$\pm$9.9\% (54) & 58.1$\pm$9.1\% (43) \\
           between & 38.8$\pm$9.7\% (67) & 22.9$\pm$9.7\% (48) & 6.7$\pm$0.0\% (15) \\
              once & 57.6$\pm$10.0\% (59) & 58.6$\pm$10.0\% (29) & 41.9$\pm$0.0\% (31) \\
             prior & 71.7$\pm$9.8\% (53) & 66.7$\pm$0.0\% (6) &  0.0$\pm$0.0\% (4) \\
            secur* & 75.8$\pm$9.5\% (91) &              - (0) &              - (0) \\
            better & 57.1$\pm$0.0\% (28) & 100.0$\pm$0.0\% (1) & 100.0$\pm$0.0\% (3) \\
              best & 27.5$\pm$9.7\% (40) & 40.0$\pm$0.0\% (5) &  0.0$\pm$0.0\% (2) \\
\hline

\end{tabularx}

\caption{The ability of syntactic patterns to reveal directives. 
The percentages result from our manual analysis, the error is given at 95\% confidence level. 
An error margin of 0\% means that we analyzed all occurrences. The number in brackets represents the number of comments we analyzed.
A dash means that the syntactic pattern can not be found in this dataset}

\label{tab:precision}

\end{table}

\subsection{Precision of Syntactic Patterns to Find Directives}
\label{precision}

As previously described, we searched the API documentation for directives by means of syntactic patterns and our tool automatically records the links between patterns and directive kinds. Hence, we are able to compute the precision of patterns -- i.e., the ratio of occurrences revealing an API directive.

 For instance, the syntactic pattern ``invo*'' occurs 50 times in the API documentation of Apache Commons Collection. If we then analyze 35 occurrences and classify 22 of them as directives, the precision of ``invo*'' would be $22/35=62.9\%$.
As discussed in Section \ref{sampling}, we can also measure the error margin of the estimation, based on the chosen  confidence level and the population size. If one analyzes all occurrences, the error margin is 0\%. In the previous example, for 35 analyzed occurrences out of 50 occurrences, the precision is $62.9\pm 9.2\%$.

Table \ref{tab:precision} presents the precision of all patterns in relation to the respective API. The numbers in brackets represent the absolute number of analyzed pattern occurrences.
We can see that the precision of patterns varies largely. 
For instance, the pattern ``between'', that we thought would identify object protocol directives, turns out to have a low precision across all APIs, i.e., many occurrences are not directives (e.g.\ \emph{``This font field editor implements chaining between a source preference store and a target preference store.''}). 
On the contrary, certain patterns have a high precision. For example, the precision of ``must'' varies between $78.6\%$ for the JDK and $98.2\%$ for Apache Commons Collections.

Furthermore, we can see that the precision of a syntactic pattern depends on the particular API.
For instance, ``warn*'' has a precision of $\approx 26.9\%$ for Java but a precision of $81.8\%$ for Commons Collections. One explanation for such huge differences is our observation that in an API's documentation often specific conventions and idioms are used. If such a convention describes a directive and contains a specific pattern, it will then result in a high precision.
Furthermore, also according to our experience, API documentation is full of redundancy and copy/pasted content. If a pattern marks such content, it also triggers a high precision. For instance \emph{``WARNING: This method is binary incompatible with Commons Collections 2.1 and 2.1.1.''} appears many times in Commons Collections, and consequently yields a high precision for ``warn*''

Given these results, we think it is neither meaningful to aggregate the precision of a syntactic pattern w.r.t.\ finding directives across different APIs nor to generalize on a syntactic pattern's precision. The precision of patterns to find directives depends too much on the application domain, on the documentation conventions, and also on the personal style of the particular documentation writers.

\subsection{Abundance of Directive Kinds}
\label{abundance}

We now present our measures of abundance for each directive kind. 

As in case of the precision, the classification of each directive's abundance may change if the dataset or the syntactic patterns are changed (cf. Section \ref{methodology}).
Table \ref{tab:abundance} summarizes our abundance measures.
The first column is the directive kind.
The second to forth columns give the absolute numbers of directives for a given directive kind and API. The fifth column summarizes the previous three columns.
The sixth column gives the frequency of this directive with respect to the number of analyzed API documentation elements (in our case $4561$).
For instance, $13\%$ of the analyzed API elements contain at least a Not Null Directive.
Finally, the column ``Abundance'' gives a quick intuition of the abundance, which is easier to grasp than the presented metric.
It is inspired from ecology, and uses the ACFOR scale (Abundant, Common, Frequent, Occasional, Rare)\footnote{see \url{http://en.wikipedia.org/wiki/Abundance_(ecology)}}.

The mapping between the number of instances and the ACFOR scale is as follows.
We first computed the average of abundance frequency A ($4.07\%$).
We say that a directive kind is abundant if it appears more than $2A\%$, common if its frequency $F\geq 1.5A\%$, frequent if $F\geq A$, occasional is $F\geq 0.5*A$ and rare otherwise.
If all directive kinds are equally abundant (i.e. $F=A$ for all kinds), the abundance would be frequent for all, which is the middle of the scale.
The goal of this mapping is to give an intuitive understanding of the numbers, even though it is arbitrary. In total, $66.5\%$ of the analyzed API elements contain at least one directive of our taxonomy.

In general, the more abundant a directive kind, the more value there is in providing automatic or semi-automatic tool support for them. 
For instance, Hovemeyer presented \cite{Hovemeyer2005} an approach to identify \code{null} references as part of the FindBug tool.
According to our results, such approaches address an important, practical problem, since ``Not Null'' directives are abundant.

We chose not to provide a comparison of the frequency of directive kinds between APIs, because there is no way to identify the controlled variable responsible for the differences. It is impossible to determine whether it would be due to the domain of the API, the documentation process, the maturity of the documentation, or to the author's style.

\begin{table}[t]
\begin{small}
\begin{tabularx}{\textwidth}{|p{6.5cm} | X | X | X | X | p{1cm} | p{1.7cm} |}

\hline
Kind &                JDK &              JFace &Col &              Total  &         \% &          Abundance \\
\hline
                 Method Call Directive &                                    965 &                                    276 &                                    754 &                                   1995 &                                 43.7\% &                               Abundant \\
                  ~~Not Null Directive &                                     28 &                                     34 &                                    531 &                                    593 &                                 13.0\% &                               Abundant \\
              ~~Return Value Directive &                                    148 &                                     52 &                                     32 &                                    232 &                                  5.1\% &                               Frequent \\
    ~~Method Call Visibility Directive &                                     78 &                                     93 &                                     22 &                                    193 &                                  4.2\% &                               Frequent \\
         ~~Exception Raising Directive &                                    156 &                                     13 &                                     16 &                                    185 &                                  4.1\% &                               Frequent \\
 ~~Miscellaneous Method Call Directive &                                    144 &                                     19 &                                     17 &                                    180 &                                  3.9\% &                             Occasional \\
            ~~Null Semantics Directive &                                     61 &                                     19 &                                     75 &                                    155 &                                  3.4\% &                             Occasional \\
             ~~String Format Directive &                                    122 &                                      1 &                                      0 &                                    123 &                                  2.7\% &                             Occasional \\
              ~~Number Range Directive &                                     80 &                                     13 &                                     29 &                                    122 &                                  2.7\% &                             Occasional \\
     ~~Method Parameter Type Directive &                                     59 &                                     10 &                                     19 &                                     88 &                                  1.9\% &                                   Rare \\
~~Method Parameter Correlation Directive &                                     75 &                                      2 &                                      8 &                                     85 &                                  1.9\% &                                   Rare \\
                 ~~Post-Call Directive &                                     14 &                                     20 &                                      5 &                                     39 &                                  0.9\% &                                   Rare \\
                 Subclassing Directive &                                    376 &                                    683 &                                     70 &                                   1129 &                                 24.8\% &                               Abundant \\
         ~~Method Overriding Directive &                                     95 &                                    364 &                                     46 &                                    505 &                                 11.1\% &                               Abundant \\
~~Extensible Class Identification Directive &                                     10 &                                    186 &                                      5 &                                    201 &                                  4.4\% &                               Frequent \\
     ~~Method Implementation Directive &                                    127 &                                     12 &                                     10 &                                    149 &                                  3.3\% &                             Occasional \\
          ~~Method Extension Directive &                                     46 &                                     45 &                                      1 &                                     92 &                                  2.0\% &                                   Rare \\
~~Non-local Consistency Subclassing Directive &                                     45 &                                     30 &                                      4 &                                     79 &                                  1.7\% &                                   Rare \\
 ~~Call Contract Subclassing Directive &                                     22 &                                     32 &                                      4 &                                     58 &                                  1.3\% &                                   Rare \\
 ~~Miscellaneous Subclassing Directive &                                     31 &                                     14 &                                      0 &                                     45 &                                  1.0\% &                                   Rare \\
                       State Directive &                                    224 &                                     92 &                                     78 &                                    394 &                                  8.6\% &                               Abundant \\
      ~~Method Call Sequence Directive &                                    157 &                                     77 &                                     50 &                                    284 &                                  6.2\% &                                 Common \\
      ~~Non Call-based State Directive &                                     67 &                                     15 &                                     28 &                                    110 &                                  2.4\% &                             Occasional \\
                 Alternative Directive &                                    199 &                                    119 &                                     75 &                                    393 &                                  8.6\% &                               Abundant \\
             Synchronization Directive &                                    105 &                                     10 &                                     61 &                                    176 &                                  3.9\% &                             Occasional \\
               Miscellaneous Directive &                                    126 &                                     13 &                                     27 &                                    166 &                                  3.6\% &                             Occasional \\
\hline
                                 Total &                                   1995 &                                   1193 &                                   1065 &                                   4253 &                                100.0\% &                                        \\
\hline
\end{tabularx} 
\end{small}

\caption{Absolute number of directives per dataset and aggregated results as percentage and abundance following the ACFOR scale. The percentage is the result of $Total\ /\ Number\ of\ analyzed\ API\ elements$ (4561). For instance, 13\% of analyzed API elements contain at least one Not Null Directive.}
\label{tab:abundance}

\end{table}

\subsection{Comparison with Dekel's Taxonomy}
\label{comparison-dekel}

Dekel \cite{Dekel2009} defines nine kinds of directives: restrictions, alternatives, protocols, parameters, return values, threading, side-effects, limitations and performance. These nine kinds of directives and ours are related as described next.

Dekel's restrictions directives are related to all kinds of method invocation restrictions. In our taxonomy, restriction is a cross-cutting concern, and we classify the directives in this category corresponding to their semantics. For instance, we classify the directive ``Subclasses should not call the super implementation.''\footnote{In JFace OwnerDrawLabelProvider.erase(event, element)} as a method extension directive, and ``It should not be called from application code.''\footnote{in JDK's Logger.setParent(parent)} as a method visibility directive.  Furthermore, we sometimes map Dekel's restriction directives to state directives and method call sequence directives in particular.

In case of the \emph{Alternative}, \emph{Protocol}, \emph{Parameter}, \emph{Return Values},  \emph{Threading} directive kinds, we basically have the same understanding and use the same category, but often identified many more concrete directives. An exception is, however, Dekel's \emph{protocols directives} which we classify as  \emph{state directives}, because -- from our point-of-view -- a protocol encodes state transitions.  Dekel's \emph{Threading directives} are called \emph{synchronization directives} in our case.

The categories ``side-effects'', ``limitations'' and ``performance'' defined by Dekel highlight a minor difference in the interpretation of the term directive between Dekel and us. For us, a directive always has an impact on the client code (e.g.\ adding tests before a method call, or the synchronized modifier in a method signature). Dekel also writes \emph{``the most important property of a directive clause is that it demands an action from the caller or suggests such an action.''}. However, to us a large part of directives from the ``side-effects'', ``limitations'' and ``performance'' categories are more warnings than directives. Contrary to directives, warnings do not imply actions from the client perspective: they highlight an implementation details or a counter-intuitive fact. This said, the frontier between directives and warnings is fuzzy and sometimes subjective, that's why we think that this difference is minor. Concretely, to us directives from the ``side-effects'', ``limitations'' and ``performance'' categories are either not directives or alternative directives when alternative solutions are given.

Also, Dekel makes a strong difference between imperative and informative directive. The former are strong requirements while the latter are more suggestions. 
We agree with this difference, but we go further: we think that all directives can have different degrees of importance. However, they are too subjective to be integrated into the taxonomy itself.
However, when Dekel statically classified each directive kind as either imperative or informative, we think that all directive kinds can be imperative, informative or somewhere in the middle.
For instance, synchronization directives, which are considered as informative by Dekel, are sometimes imperative to our opinion (e.g. \emph{this call has to happen in a synchronized block}).
Hence, our taxonomy has no such dichotomy.

Finally, when compared to Dekel's work, our subclassing related directives are more fine-grained and differentiated. 
Dekel only mentions return value directives while we identified and discussed nine other directive kinds based on subclassing.

\subsection{Guidelines on Writing API Directives}

\label{guidelines}

This section presents general guidelines on how to write good API directives.
Those guidelines emerged from our experience of reading thousands of API comments as part of our empirical study.\footnote{
We exclusively focus on API directives and not on other concerns of API documentation (audience, writing process, etc.). For a comprehensive discussion of these other concerns, we refer to the excellent paper written by a senior technical writer at Sun: ``API documentation from source code comments: a case study of Javadoc'' \cite{kramer1999api}.}

\subsubsection*{Precisely identify the target API elements of a directive} 
An API directive can refer to any API element and in particular may refer to other elements than the one where it is documented. For example, a directive that is stated in a class' documentation may refer to the usage of the class, but also to other classes with which the class interacts, or the methods and fields. In this case, it is very important that the concerned API elements are precisely identified. To avoid ambiguous references consider explicitly naming the related API elements.

\subsubsection*{Precisely specify when a directive applies} 
Many API directives exist that only apply if a certain condition is met, i.e.\ they do not state absolute requirements such as \emph{``this method parameter must not be null''}. Examples of such directives include \emph{``Subclasses may override this method if ...''} or \emph{``Instead use method X if ...''}. It is very important to always clearly state under which conditions a directive applies. Otherwise, developers must either guess them or will just ignore the directive.

\subsubsection*{Avoid ``should''}
Most directives directly affect the client code. Hence, it is necessary that a directive is precise and that it is not vague. Otherwise, the developer has to write code to accommodate for cases that may never occur.  For example, the directive ``this method should not return null'' indicates that the method will probably never return \code{null}, but this is not certain. In general, during our case-study, we found that when the syntactic patterns ``should'' and ``should not'' are used, they raise more questions than they clarify correct usage.

\subsubsection*{Decide and document the terminology}
We have experienced that there are sometimes different usages of certain important words or expressions (e.g. ``Anti-pattern'' in Section \ref{method-overriding-directives}). In order to maximize the clarity of directives, it is important to find a consensus on a fixed terminology (words and co-locations) within the library development team, to document it and to extensively refer to the terminology document in the API documentation.

\section{Conclusion and Future Work}
\label{conclusion}

In this paper, we discussed the protocol and the results of an exploratory case-study that we carried out to determine the nature of directives in API documentation. 
We presented and discussed a comprehensive taxonomy of API directives.
The taxonomy was constructed by analyzing more than 4000+ API documentation items. We found that $66.5\%$ of the analyzed API elements contain at least one directive from our taxonomy.
All directive kinds that we found were described using a rigorous template to facilitate comprehension, comparison and future use.
Most of the directive kinds of the taxonomy were either not precisely described or even not identified in previous work.

Given the rigor and scope of the performed analysis of API directives, this work aims to serve as an empirical foundation for future work on related software engineering tools. Consequently, the extracted directive data is available as electronic  supplementary material and in \cite{paperdata}. For example, it can serve as the foundation for the development of next-generation tools for the automated extraction of API directives from documentation (see \cite{Zhong2009} for a seminal paper on this topic). For such tools, this work provides the necessary information about the kind of directives to look for and also information about the syntactic patterns that are actually used. 

A second set of tools -- for which this work can serve as a foundation -- are those related to the formal specification and checking of API directives. As discussed in, e.g., \cite{Nanda2009}, certain kinds of directives can be expressed in a machine-processable manner  such that automatic checking of client code is possible. 
For instance, one can imagine new Javadoc annotations such as \texttt{@NumberRange}.
In this case, this work helps to identify those directives that are actually used and, hence, for which kinds of directives it is meaningful to provide corresponding support. One concrete example in this area is Eclipse's \code{@noinstantiate} annotation. 

A third set of tools for which the presented results are important are those that try to infer API directives. Various directives could directly be inferred from the API's source code using static analysis (e.g. \cite{Thummalapenta2008} for a small subset of subclassing directives) or from client code (e.g. \cite{Stylos2009a}). For instance, code such as \emph{ if (param == null) \{ throw new IllegalArgumentException(); \}} at the beginning of a method probably implies a ``Not Null'' directive. While the fields of static and dynamic analysis already contain related contributions, there is still room for devising a complete tool chain from the automated inference of formal models to generating natural language API documentation targeting human developers.

\bibliographystyle{abbrv}

\bibliography{article}

\end{document}